\documentclass[a4paper,12pt]{article}
\usepackage{graphicx}

\begin{document}

\begin{center}
October, 2005\\
\begin{large}
{\bfseries Tunnel transport through multiple junctions}\\
\vspace{.4cm}
J. Peralta-Ramos$^{a,*}$ and A. M. Llois$^a$\\
\end{large}
\vspace{0.2cm}
\begin{small}
{\itshape
$^a$ Departamento de F\'isica, Centro At\'omico Constituyentes, Comisi\'on 
Nacional de Energ\'ia At\'omica}\\
\end{small}

\end{center}
\vspace{1cm}

\begin{small}
We calculate the conductance through double junctions of the type 
M(inf.)-Sn-Mm-Sn-M(inf.) and triple junctions of the type 
M(inf.)-Sn-Mm-Sn-Mm-Sn-M(inf.), where M(inf.) are semi-infinite metallic 
electrodes, Sn are 'n' layers of semiconductor and Mm are 'm' layers of metal 
(the same as the electrodes), and compare the results with the conductance 
through simple junctions of the type M(inf.)-Sn-M(inf.). The junctions are 
bi-dimensional and their parts (electrodes and 'active region') are periodic in the 
direction perpendicular to the transport direction. To calculate the conductance 
we use the Green's Functions Landauer-B$\ddot{u}$ttiker formalism. The electronic 
structure of the junction is modeled by a tight binding Hamiltonian.\\ 
For a simple junction we find that the conductance decays exponentially with 
semiconductor thickness. For double and triple junctions, the conductance 
oscillates with the metal in-between thickness, and presents peaks for which the 
conductance is enhanced by 1-4 orders of magnitude. We find that when 
there is a conductance peak, the conductance is higher to that corresponding to 
a simple junction. The maximum ratio 
between the conductance of a double junction and the conductance of a 
simple junction is 146 $\%$, while for a triple junction it is 323 
$\%$. These oscillations in the conductance are explained in terms of the energy
spectrum of the junction's active region.
\vspace{0.3cm}
$^{*}$ Corresponding author

Centro At\'omico Constituyentes, Av. Gral. Paz 1499, San Mart\'in (1650), 
Buenos Aires, Argentina

TEL: (54-11) 6772-7007 FAX: (54-11) 6772-7121

peralta@cnea.gov.ar
\end{small}

\vspace{1cm}
A magnetic tunnel junction (MTJ) consists of a few layers of a non-conducting 
material placed between two ferromagnetic electrodes. If we apply a voltage to 
the electrodes, a current will flow through the junction. It is observed 
experimentally that the magnitude of this current depends on the relative 
orientation of the magnetization in each electrode. This phenomenon is termed 
tunneling magnetoresistance (TMR), and is nowadays being extensively studied 
since it forms the basis of what is called 'spintronics'.\\
Simple MTJ's like that described above have been extensively studied [1], 
but double and triple junctions, in which metallic layers are inserted between 
the non-conducting material, have not. Some recent conductance measurements in 
double junctions [2,3] show a large increase in the conductance as 
compared to the conductance of simple junctions composed of the same materials, 
indicating that multiple junctions may present an advantage over simple 
ones.
The aim of this contribution is to study, within a very simple model for the 
junction's electronic structure, what happens with the conductance of a simple 
junction when we insert metallic layers in between the semiconductor.\\
As a model system, we consider a semiconductor whose 
structure is given by a square Bravais lattice of constant $a=3.2$ \AA ~with two 
atoms per unit cell, A and B. Each semiconductor layer consists of an 
infinite array of atoms A and B in the $y$ direction, and has a thickness of
$3.2$ \AA~, as shown in Fig. 1. In a 
simple junction, the 
semiconductor slab is 
sandwiched by two semi-infinite metallic electrodes of the same structure
which are 
paramagnetic, composed of atoms C and D instead of A and B. Epitaxial growth of 
the spacer on the metallic leads is assumed. The electronic structure of the 
junction is modeled by a tight-binding Hamiltonian with one $s$ orbital per 
site and second nearest neighbors interaction. For the semiconductor, the on-site 
energies are $E_A=1$ eV and $E_B=1.5$ eV, and the hopping parameter is 
$t=0.2$ eV for first and second nearest-neighbors. With these 
parameters, a band gap of $0.5$ eV is obtained. For the 
electrodes, the parameters are $E_C=2.8$ eV, $E_D=2.9$ eV and
$t=1$ eV. The Fermi energy $E_F$ falls in the middle 
of the spacer's band gap. To construct double and triple junctions, metallic layers
identical to the electrodes 
are inserted in between the semiconductor layers of a simple junction.
Fig. 2 shows schematically the structure of simple, 
double and triple junctions. The active region consists of a 'semiconductor region' 
and in-between metallic layers.\\ 
The conductance is calculated using the Landauer-B$\ddot{u}$ttiker formalism
[4] expressed in terms of the active region's Green's function 
$G_S=[\hat{1}E-H_S-\Sigma_L-\Sigma_R]^{-1}$, where $\hat{1}$ stands for the unit matrix, 
$H_S$ is the Hamiltonian corresponding to the active part of the 
junction, and $\Sigma_{L/R}$ are the self-energies 
describing the interaction of the active region with the left (L) or right (R) 
electrodes. In this expression, $E$ is actually $E+i\eta$, where $\eta>0$ is 
a very small real number. This Green's function describes the propagation of
an electron through the active region {\bfseries taking into account, via the 
self-energies, the presence of the electrodes}.
The self-energies are given by 
$\Sigma_L=H_{LS}^{\dagger} g_L H_{LS}$ and $\Sigma_R=H_{RS}^{\dagger} g_R 
H_{RS}$, where $H_{LS}$ and $H_{RS}$ describe the coupling of the active
region with 
the electrodes, and $g_{L/R}$ are the surface Green's functions for each 
electrode. These surface Green's functions are calculated using a semi-analytical 
method [5] and are exact within our tight-binding approximation. The 
transmission probability $T$ is given by [4]
$T(k_{y},E)=Tr~ [\Delta_L G_S \Delta_R G_S ^{\dagger}]$
where $\Delta_{L/R}=i (\Sigma_{L/R}-\Sigma_{L/R}^{\dagger})$, while
the conductance is given by 
\begin{equation}
\Gamma(E)=\frac{2e^2}{h}\frac{1}{N_{k_{y}}} \sum_{k_y} T(k_y,E)
\end{equation}
where the $2$ comes from spin degeneracy and $N_{k_{y}}$ is the total number of
wave vectors considered (in our case 400 is enough to achieve stable values of 
$\Gamma$). In this work we restrict to $E=E_F$.\\
For a simple junction, it is found that the conductance decays exponentially 
with semiconductor thickness, with a decay parameter in excellent agreement 
with its complex band structure [6,7]. For double and triple junctions, 
the conductance presents peaks for certain thicknesses of the in-between metal, 
in which the conductance raises by 1 to 4 orders of magnitude with respect to
other thicknesses. The peak
conductance is more than one order of magnitude larger for a multiple junction
as compared to the conductance of a simple junction, and this conductance is
higher for triple than for double junctions. 
Fig. 3 shows 
as an example the conductance of a double junction of 2 semiconductor layers
as a function of the number of layers of the metal in between. It is found that
for 2, 6 and 9 layers of the metal in between there are peaks in the conductance.
The ratio between the maximum conductance of a double or triple junction and the
conductance of a simple junction is shown in Fig. 4. It is seen that the maximum
ratio is 146 $\%$ for a double junction and 323 $\%$ for a triple junction, which
represent very large increases in the conductance of these multiple junctions. 
For a double junction, the largest increase occurs for 3 semiconductor layers 
at each side of
the metal layers, while for a triple junction it occurs for 4 semiconductor layers.
Lastly, for those in-between metal thicknesses for which there is a conductance peak,
it is found that the energy spectrum of the junction's active
region possesses an eigenenergy very close to the Fermi 
energy of the electrodes, and that this does not happen for other in-between metal
thicknesses. As confirmed by density of states calculations, the insertion 
of metal 
layers in between the semiconductor results in
the appearance of electronic states with energies in the semiconductor's band gap. 
By calculating partial densities of states projected onto each active region's atom, 
it is 
found that these states are not localized, and, to the contrary, extend throughout
the whole active region.
For certain thicknesses of the in-between metallic layers, some of these 
band gap states have an energy very close to $E_F$, thus producing a resonance state
which extends throughout the whole junction, including the electrodes.\\
In conclusion, by using a very simple model junction we have made
plausible a situation in which the conductance
of a simple junction largely increases by the insertion of metal in between the 
non-conducting material, and have also shown that this effect is to be traced back
to resonance states that can
extend throughout the whole junction. We believe
that by making the electrodes ferromagnetic, the conductance peaks for each spin channel
will occur at different thicknesses of the in-between metal, thus allowing to tune
the values of TMR (the ratio between the conductance in the parallel and
antiparallel configuration) by varying this thickness. These issues are 
currently being investigated.\\
This work was partially funded by UBACyT-X115, 
Fundaci\'on Antorchas and PICT 03-10698. Ana Mar\'ia Llois belongs to CONICET 
(Argentina). 

\begin{center}
References\\
\end{center}
$[1]$ X-G Zhang {\itshape et al}, J. Phys.: Condens. Matter 
{\bfseries 15} (2003) R1603-R1639\\
$[2]$ J. H. Lee {\itshape et al}, J. Magn. Magn. Mater. {\bfseries 286} (2005),
138-141\\
$[3]$ T. Nozaki {\itshape et al}, Appl. Phys. Lett. {\bfseries 86}, 082501 (2005)\\
$[4]$ S. Datta, 'Electronic transport in mesoscopic systems',
Cambridge University Press, United Kingdom, 1999\\
$[5]$ S. Sanvito {\itshape et al}, Phys. Rev. B {\bfseries 59},
11936 (1999)\\
$[6]$ Ph. Mavropoulos {\itshape et al}, Phys. Rev. Lett. 
{\bfseries 85} (2000) 1088\\
$[7]$ Jer\'onimo Peralta-Ramos {\itshape et al}, Physica B {\bfseries 354}
(2004) 166-170\\


Figure 1\\
Structure of the semiconductor. The system is periodic in the 
$y$ direction. The electrodes
are semi-infinite in the $x$ direction and have the same structure 
as the semiconductor.\\

Figure 2\\
Schematic structure of a simple, double and triple junction.\\

Figure 3\\
Conductance at the Fermi energy as a function of the number of layers of
the in-between metal, for a double junction with 2 semiconductor layers at
each side of the in-between metal.\\

Figure 4\\
Ratio between the largest conductances obtained for double and
triple junctions and that of
the simple junction, as a function of the number of layers in the semiconducting
regions. 

\newpage


\begin{figure}
\scalebox{0.5}{\includegraphics{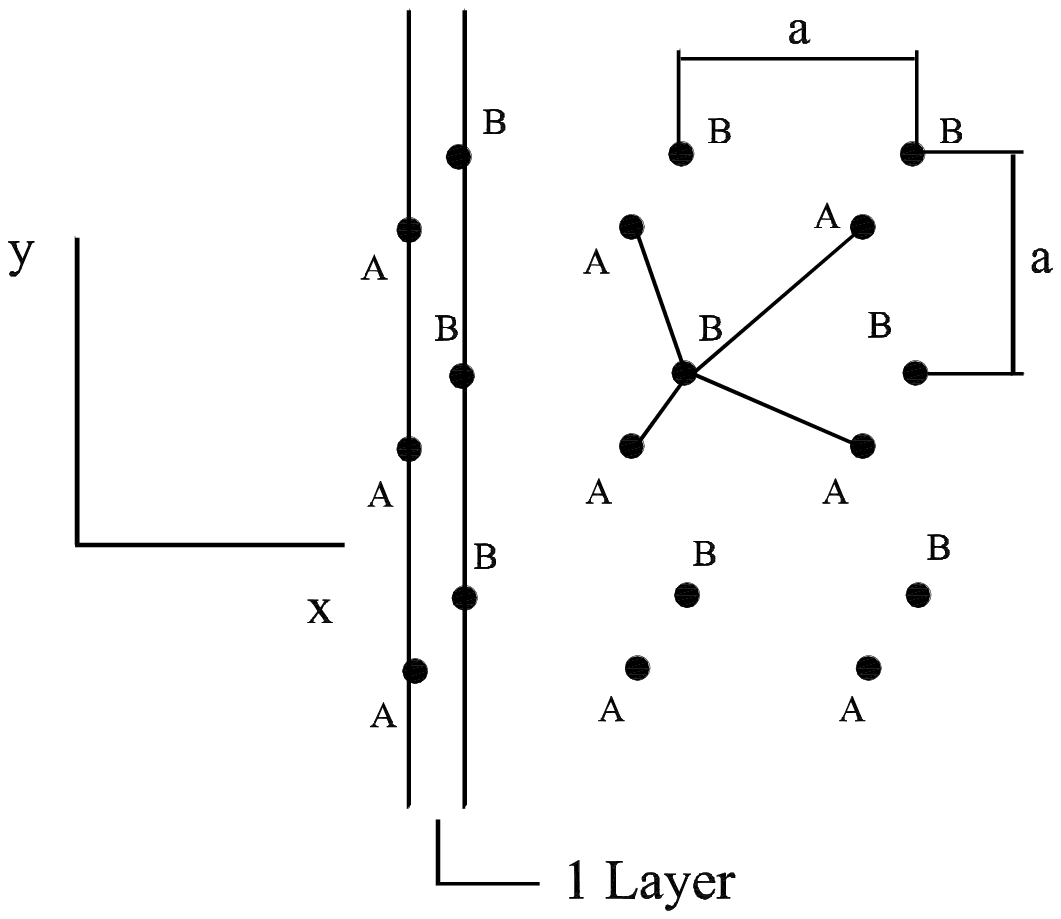}}
\caption{}
\end{figure}

\begin{figure}
\scalebox{0.7}{\includegraphics{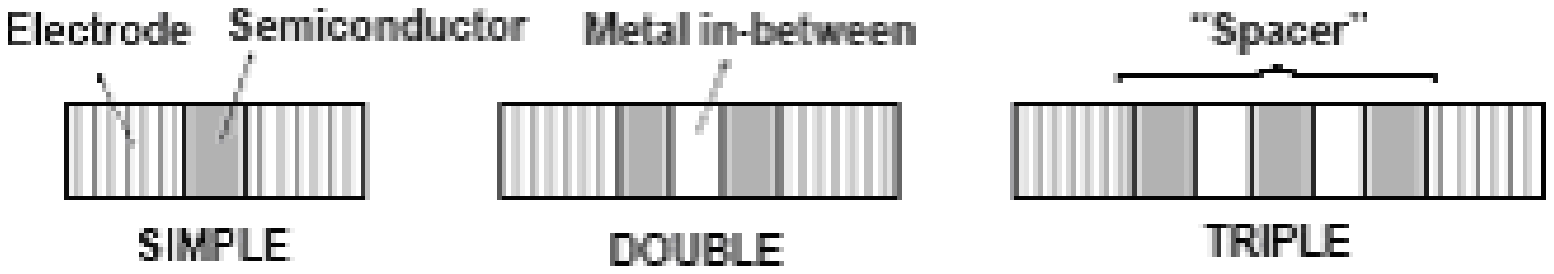}}
\caption{}
\end{figure}

\begin{figure}
\scalebox{0.6}{\includegraphics{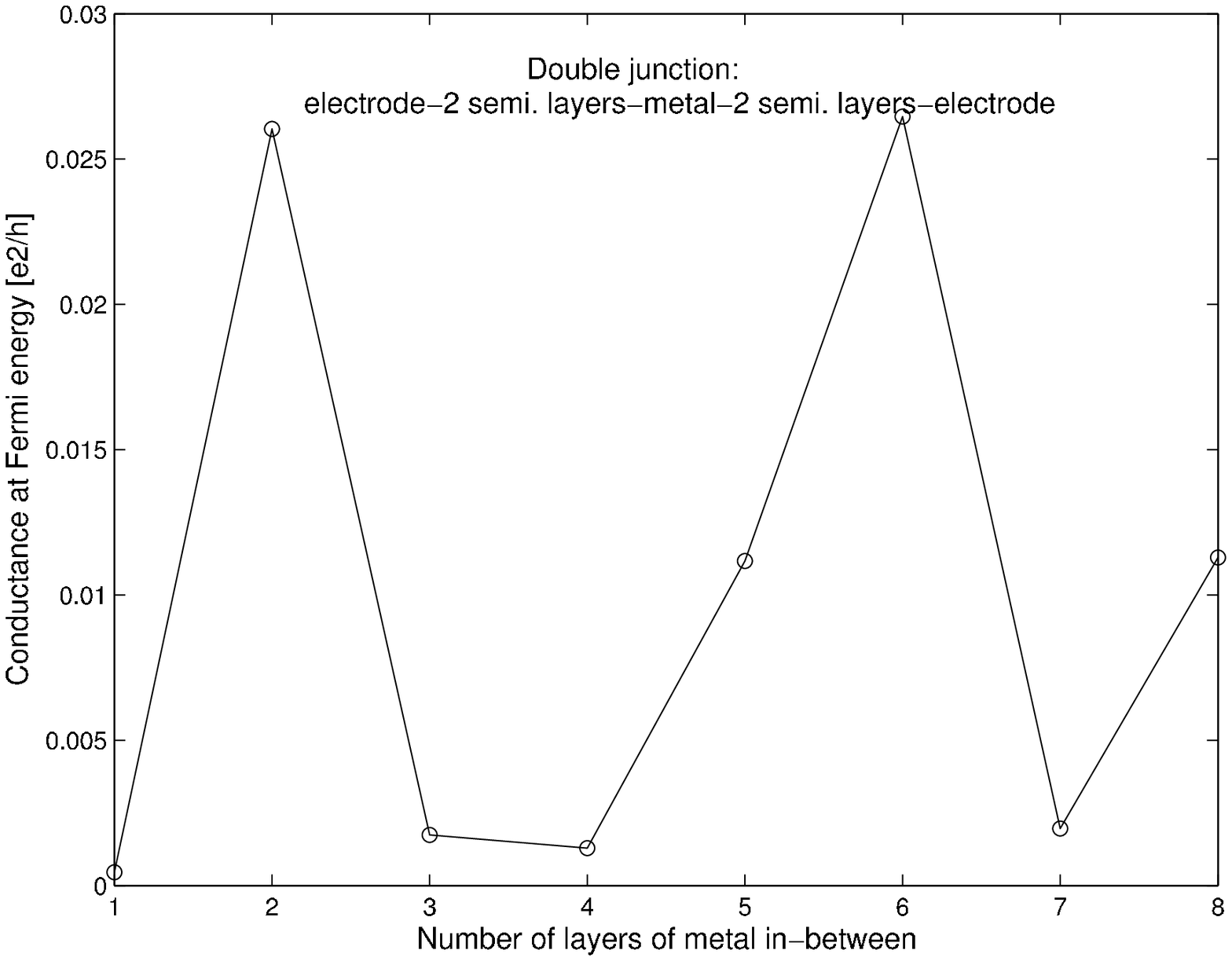}}
\caption{}
\end{figure}

\begin{figure}
\scalebox{0.6}{\includegraphics{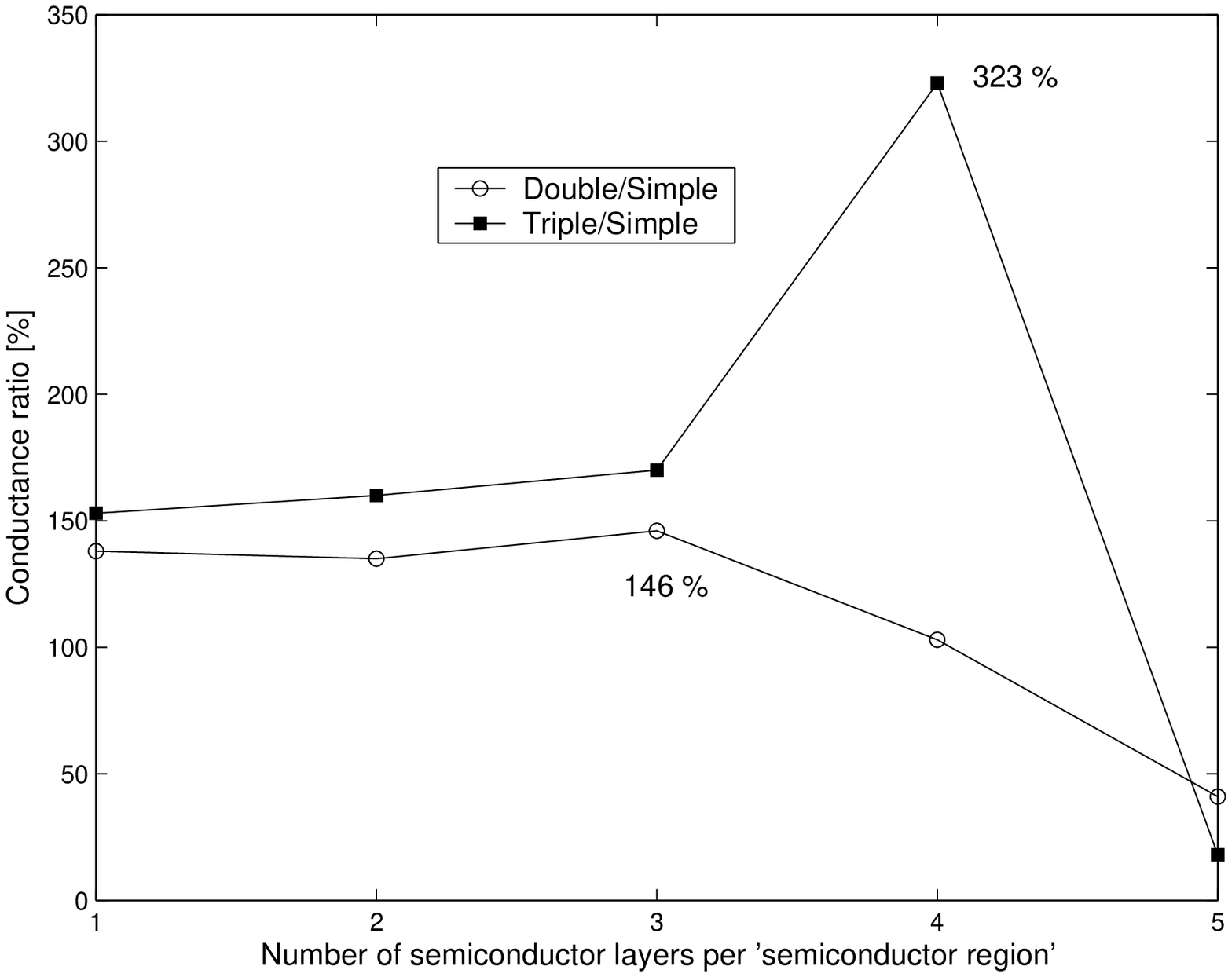}}
\caption{}
\end{figure}

\end{document}